\documentclass{reporter}

\usepackage[T1]{fontenc}
\usepackage[tracking]{microtype}
\usepackage{libertine}   
\usepackage[euler-digits,small]{eulervm}
\AtBeginDocument{\renewcommand{\hbar}{\hslash}}

\usepackage{siunitx}
\usepackage[xindy]{glossaries}
\usepackage{tikz}
\usepackage{color}
\usepackage{pifont}
\usepackage{natbib}
\usepackage[hidelinks, citecolor=red]{hyperref}

\definecolor{blue}{HTML}{348ABD}
\definecolor{purple}{HTML}{7A68A6}
\definecolor{red}{HTML}{aa0000}
\definecolor{green}{HTML}{408040}
\definecolor{orange}{HTML}{E24A33}

\newacronym{bss}{BSS}{BrainScaleS Project}
\newacronym{hbp}{HBP}{Human Brain Project}
\newacronym{msb}{MSB}{Most Significant Bit}
\newacronym{lsb}{LSB}{Least Significant Bit}
\newacronym{adex}{AdEx}{Adaptive Exponential Integrate-and-Fire model}
\newacronym{stdp}{STDP}{Spike-Timing Dependent Plasticity}
\newacronym{stp}{STP}{Short Term Plasticity}
\newacronym{stdf}{STDF}{Short Term Depression and Facilitation}
\newacronym{psp}{PSP}{Postsynaptic Potential}
\newacronym{epsp}{EPSP}{Excitatory Postsynaptic Potential}
\newacronym{hicann}{HICANN}{High Input Count Analog Neural Network}
\newacronym{hmf}{HMF}{Hybrid Multi-Scale Facility}
\newacronym{fpga}{FPGA}{Field Programmable Gate Array}
\newacronym{adc}{ADC}{Analog Digital Converter}
\newacronym{htd}{HTD}{Hardware Time Domain}
\newacronym{btd}{BTD}{Biological Time Domain}
\newacronym{vlsi}{VLSI}{Very Large Scale Integration}
\newacronym{asic}{ASIC}{Application Specific Integrated Circuit}
\newacronym{lif}{LIF}{Leaky Integrate-and-Fire}
\newacronym{kwta}{kWTA}{$k$-Winners-Take-All}
\newacronym{htm}{HTM}{Hierarchical Temporal Memory}
\newacronym{sdr}{SDR}{Sparse Distributed Representation}
\newacronym{ppu}{PPU}{Plasticity Processing Unit}

\makeglossaries

\newcommand*\circled[2]{\tikz[baseline=(char.base)]{
	\node[shape=circle,fill=white,draw=#1,inner sep=1pt] (char) {\color{white}\textbf{#2}};
	\node[shape=circle,fill=#1,inner sep=0.5pt] (char) {\color{white}\textbf{#2}};
	}}

\begin{document}
\firstpage{1}

\title[Porting HTM Models to the Heidelberg Neuromorphic Computing Platform]{Porting HTM Models to the Heidelberg Neuromorphic Computing Platform}
\author[Billaudelle et al.]{Sebastian Billaudelle\,$^{1,}$\footnote{email: \href{mailto:sebastian.billaudelle@kip.uni-heidelberg.de}{sebastian.billaudelle@kip.uni-heidelberg.de}}\ , Subutai Ahmad\,$^{2,}$\footnote{email: \href{mailto:sahmad@numenta.com}{sahmad@numenta.com}}}
\address{$^{1}$Kirchhoff-Institute for Physics, Heidelberg, Germany\\
$^{2}$Numenta, Inc., Redwood City, CA}

\history{Received on XXXXX; revised on XXXXX; accepted on XXXXX}

\editor{Associate Editor: XXXXXXX}

\maketitle

\begin{abstract}

\gls{htm} is a computational theory of machine
intelligence based on a detailed study of the neocortex. The Heidelberg
Neuromorphic Computing Platform, developed as part of the \gls{hbp},
is a mixed-signal (analog and digital) large-scale platform for
modeling networks of spiking neurons. In this paper we present the first effort
in porting \gls{htm} networks to this platform.  We describe a framework for
simulating key \gls{htm} operations  using spiking network models. We then describe
specific spatial pooling and temporal memory implementations, as well as
simulations demonstrating that the fundamental properties are maintained. We
discuss issues in implementing the full set of plasticity rules using \gls{stdp}, and
rough place and route calculations. Although further work is required, our
initial studies indicate that it should be  possible to run large-scale \gls{htm}
networks (including plasticity rules) efficiently on the Heidelberg platform.
More generally the exercise of porting high level \gls{htm} algorithms to biophysical
neuron models promises to be a fruitful area of investigation for future
studies.

\end{abstract}

\glsresetall

\section{Introduction}
The mammalian brain, particularly that of humans, is able to process diverse
sensory input, learn and recognize complex spatial and temporal patterns, and
generate behaviour based on context and previous experiences. While computers
are efficient in carrying out numerical calculations, they fall short in solving
cognitive tasks. Studying the brain and the neocortex in particular is an
important step to develop new algorithms closing the gap between intelligent
organisms and artificial systems. Numenta is a company dedicated to developing
such algorithms and at the same time investigating the principles of the
neocortex. Their \gls{htm} models are designed to solve real world problems
based on neuroscience results and theories.

Efficiently simulating large-scale neural networks in software is still a
challenge. The more biophysical details a model features, the more computational
ressources it requires. Different techniques for speeding up the execution of
such implementations exist, e.g. by parallelizing calculations. Dedicated
hardware platforms are also being developed. Digital neuromorphic hardware like
the SpiNNaker platform often features highly parallelized processing
architectures and optimized signal routing \citep{furber2014spinnaker}. On the
other hand, analog systems directly emulate the neuron's behavior in electronic
microcircuits. The \gls{hmf} is a mixed-signal platform developed in the scopes
of the \gls{bss} and \gls{hbp}.

In this paper we present efforts in porting \gls{htm} networks to the \gls{hmf}.
A framework for simulating \glspl{htm} based on spiking neural networks is
introduced, as well as concrete network models for the HTM concepts spatial
pooling and the temporal memory. We compare the behavior to software
implementations in order to verify basic properties of the \gls{htm} networks.
We discuss the overall applicability of these models on the target platform, the
impact of synaptic plasticity, and connection routing considerations.

\subsection{Hierarchical Temporal Memory}

\gls{htm} represents a set of concepts and algorithms for machine intelligence
based on neocortical principles \citep{numenta2011htm}. It is designed to learn
spatial as well as temporal patterns and generate predictions from
previously seen sequences. It features continuous learning and operates
on streaming data. An \gls{htm} network consists of one or multiple
hierarchically arranged regions. The latter contain neurons organized in
columns. The functional principle is captured in two algorithms which are laid
out in detail in the original whitepaper \citep{numenta2011htm}. The following
paragraphs are intended as an introductory overview and introduce the
properties relevant to this work.

\label{sec:spatial_pooler_properties}

The \emph{spatial pooler} is designed to map a binary input vector to a set of
columns. By recognizing previously seen input data, it increases stability and
reduces the system's susceptibility for noise. Its behaviour can be
characterized by the following properties:

\begin{enumerate}
	\item\label{enm:spatial_pooler_sparsity} The columnar activity is sparse.
	Typically, 40 out of 2,048 colums are active, which is approximately a
	sparsity
	of \SI{2}{\%}. The number of active columns is constant in each time step and
	does not depend on the input sparsity.

	\item\label{enm:spatial_pooler_selection} The spatial pooler activates the $k$
	columns which receive the most input. In case of a tie between two columns, the
	active column is selected randomly, e.g. through structural advantages of
	certain cells compared to its neighbors.

	\item\label{enm:spatial_pooler_overlap} Stimuli with low pairwise overlap
	counts are mapped to sparse columnar representations with low pairwise
	overlap counts, while high overlaps are projected onto representations
	with high overlap. Thus, similar input vectors lead to a similar columnar
	activation, while disjunct stimuli activate distinct columns.

	\item\label{enm:spatial_pooler_minimum} A column must receive a minimum input
	(e.g. 15 bits) to become active.

\end{enumerate}


The \emph{temporal memory} operates on single cells within columns and further
processes the spatial pooler's output. Temporal sequences are learned by the
network and can be used for generating predictions and highlighting anomalies.
Individual cells receive stimuli from other neurons on their distal dendrites.
This lateral input provides a temporal context. By modifying a cell's distal
connectivity, temporal sequences can be learned and predicted. The temporal
memory's behavior can be summarized by the following:

\begin{enumerate}

	\item Individual cells receive lateral input on their distal dendrites. In
	case a certain threshold is crossed, the cells enter a predictive
	(depolarized) state.

	\item\label{enm:temporal_memory_predictive} When a column becomes active due
	to proximal input, it activates only those cells that are in predictive state.

	\item\label{enm:temporal_memory_bursting} When a column with no predicted
	cells becomes active due to proximal input, all cells in the column become
	active. This
	phenomenon is referred to as columnar bursting.

\end{enumerate}

\subsection{Heidelberg Neuromorphic Computing Platform}

\begin{figure}
	\begin{center}
		\includegraphics[width=\columnwidth]{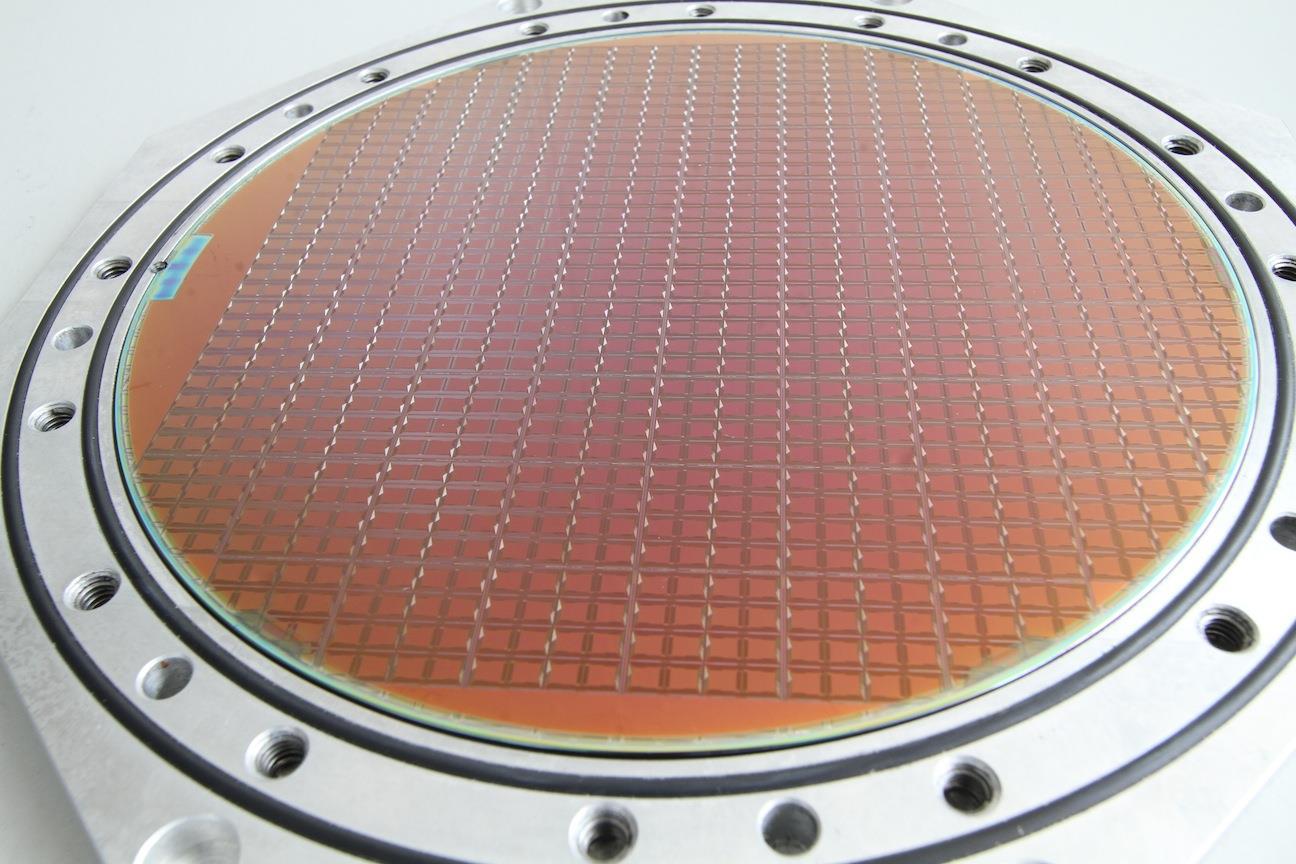}
	\end{center}
	\caption{A wafer containing 384 \gls{hicann} chips. The undiced wafer undergoes a custom post-processing step where additional metal layers are applied to establish inter-reticle connectivity and power distribution. (Photo courtesy of the Electronic Vision(s) group, Heidelberg.)}
	\label{fig:wafer}
\end{figure}

The \gls{hmf} is a hybrid platform consisting of a traditional high-performance
cluster and a neuromorphic system. It is developed primarily at the
Kirchhoff-Institute for Physics in Heidelberg and the TU Dresden while receiving
funding from the \gls{bss} and \gls{hbp} \citep{hbp2014sp9spec}. The platform's
core is the wafer-scale integrated \gls{hicann} chip as shown in
Figure~\ref{fig:wafer}. Part of the chip's unique design is its mixed-signal
architecture featuring analog neuron circuits and a digital communication
infrastructure. Due to the short intrinsic time constants of the hardware
neurons, the system operates on an accelerated timescale with a speed-up factor
of \num{10e4} compared to biological real-time.

\gls{hicann} features 512 neurons or \emph{dendritic membrane circuits}. Each
circuit can be stimulated via 226 synapses on two synaptic inputs. As a default,
the latter are configured for excitatory and inhibitory stimuli, respectively.
However, they can be set up to represent e.g. two excitatory inputs with
different synaptic time constants or reversal potentials. By connecting multiple
dendritic membranes larger neurons with up to \num{14e3} synapses can be formed.

A single wafer contains 384 chips with \num{200e3} neurons and \num{45e6} synapses. Multiple wafers can be connected to form even larger networks. The \gls{bss}'s infrastructure consists of six wafers and is being extended to 20 wafers for the first \gls{hbp} milestone.

%
%

\subsection{Spiking Neuron Model}

There exist different techniques of varying complexity for simulating networks
of spiking neurons. The reference implementation we use for \gls{htm}  networks
is  based on first generation, binary neurons with discrete time steps
\citep{nupic}. Third generation models, however, incorporate the concept of
dynamic time and implement inter-neuron communication based individual spikes.

Starting from the original Hodgkin-Huxley equations \citep{Hodgkin1952},
multiple spiking neuron models were developed that feature different levels of
detail and abstraction. The \gls{hicann} chip implements \gls{adex} neurons
\citep{brette2005adaptive}. At its core, it represents a simple \gls{lif} model
but features a detailed spiking behavior as well as spike-triggered and
sub-threshold adaption. It was found to correctly predict approximately
\SI{96}{\%} of the spike times of a Hodgkin-Huxley-type model neuron and about
\SI{90}{\%} of the spikes recorded from a cortical neuron
\citep{jolivet2008quantitative}. On the \gls{hmf} and thus also in the following
simulations, the neurons are paired with conductance-based synapses allowing for
a fine-grained control of the synaptic currents and the implementation of e.g.
shunting inhibition.

\section{Spiking Network Models}
Implementing neural network models for a neuromorphic hardware platform or
dynamic software simulations requires an abstract network description defining
the individual cell populations as well as the model's connectivity. For this
work, our primary focus was on developing mechanistic and functional
implementations of the software reference models while staying within the
topological and parameter restrictions imposed by the hardware platform.
A more detailed biophysical approach should begin with simulations of single
\gls{htm} neurons and their dendritic properties before advancing to more
complex systems, e.g. full networks.

In the following sections we describe spatial pooler and temporal memory
models that incorporate basic \gls{htm} properties. These models are able to
reproduce the fundamental behaviour of existing software implementations.

The simulations were set up in Python using the PyNN library \citep{davison2008pynn}. Besides supporting a wide range of software simulators, this high-level interface is also supported by the \gls{hmf} platform \citep{billaudelle14pyhmf}. NEST was used as a simulation backend \citep{gewaltig2007nest}. To enable multiple synaptic time constants per neuron, a custom implementation of the \gls{adex} model was written.

\subsection{Spatial Pooler}
\label{sss:spatial_pooler_network}

At its core the spatial pooler resembles a \gls{kwta}
network: $k$ out of $m$ columns are chosen to be active in each time step. In
fact, \gls{kwta} networks have often been mentioned as an approximation for
circuits naturally occurring in the neocortex \citep{felch2008hypergeometric}.
Continuous-time and VLSI implementations of such systems have been discussed
in the literature \citep{erlanson1991analog,tymoshchuk2012,maass2000neural}.
In the implementation below we describe a novel approach to maintaining
stable sparsity levels even with a large number of inputs.

\begin{figure}
	\begin{center}
		\includegraphics[width=\columnwidth]{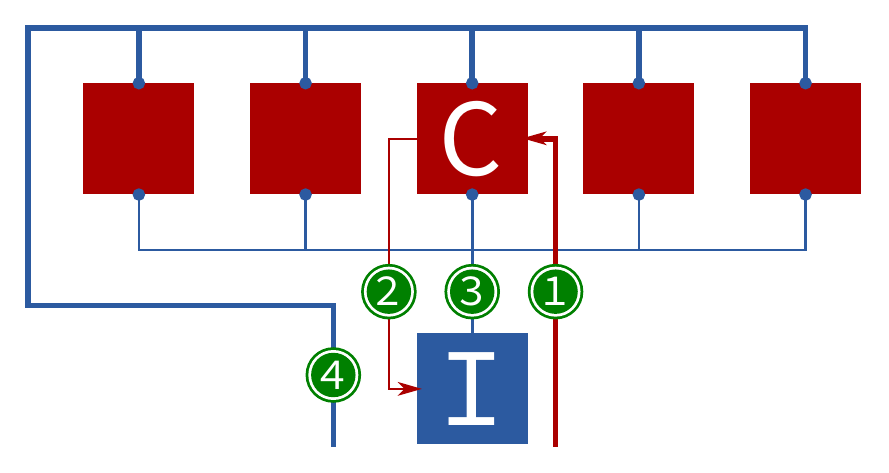}
	\end{center}
	\caption{Timing based implementation of the spatial pooler. Each column is
	represented by a single cell \emph{C} and receives sparse input from the input
	vector \protect\circled{green}{1}. The columns become active when the number of
	connected active inputs crosses a threshold. The rise time of the membrane
	voltage highly depends on the number of coincident inputs: cells with more
	presynaptic activity will fire before those with less stimuli do. Inhibitory
	pool \emph{I} accumulates the columnar spikes \protect\circled{green}{2} and in
	doing so acts as a counter. After a certain number of columns have become
	active, the pool will inhibit and shut down all columns preventing any further
	activity \protect\circled{green}{3}. To stabilize this \gls{kwta} model, all
	columns receive a subsampled feed-forward inhibition
	\protect\circled{green}{4}. This effectively prolongs the decision period for
	high input activity.}

	\label{fig:spatial_pooler}
\end{figure}

The network developed for this purpose is presented in
Figure~\ref{fig:spatial_pooler}. It follows a purely time-based approach and
is designed for \gls{lif} neurons. It allows for very fast decision processes
based on a single input event per source. Each column is represented by a single
cell which accumulates feed-forward input from the spike sources. Here, the rise
time of the membrane voltage decreases with the number of presynaptic events
seen by the cell: cells receiving the most input will fire before the others. An
inhibitory pool consisting of a single cell collects the network's activity. Low
membrane and high synaptic time constants lead to a reliable summation of
events. When a certain number of spikes have been collected -- and thus the
cell's threshold has been crossed -- the pool strongly inhibits all cells of
the network suppressing subsequent spike events.

The model is extended by adding subtle feed-forward shunting inhibition. The
inhibitory conductance increases with the overall input activity
$\nu_\text{in}$. With the reversal potential set to match the leakage potential,
the conductance contributes to the leakage term

\begin{align*}
	g_\text{l}' &= g_\text{l} + g_\text{inh}(\nu_\text{in}) \,.
\end{align*}
This effectively slows down the
neurons' responses and thus prolongs the decision period of the network. With
this inhibition, the resulting system is able to achieve stable sparsity
levels with a large number of inputs, at the cost of slightly slower response
times.

Tie situations between columns receiving the same number of presynaptic events
are resolved by adding slight gaussian jitter to the weights of the
excitatory feed-forward connections. This gives some columns structural
advantages over other columns resulting in a slightly faster response to the
same stimulus. By increasing the standard deviation $\sigma_\text{j}$ of the
jitter, the selection criterion can be blurred.


\subsection{Temporal Memory}

\begin{figure}
	\begin{center}
		\includegraphics[width=\columnwidth]{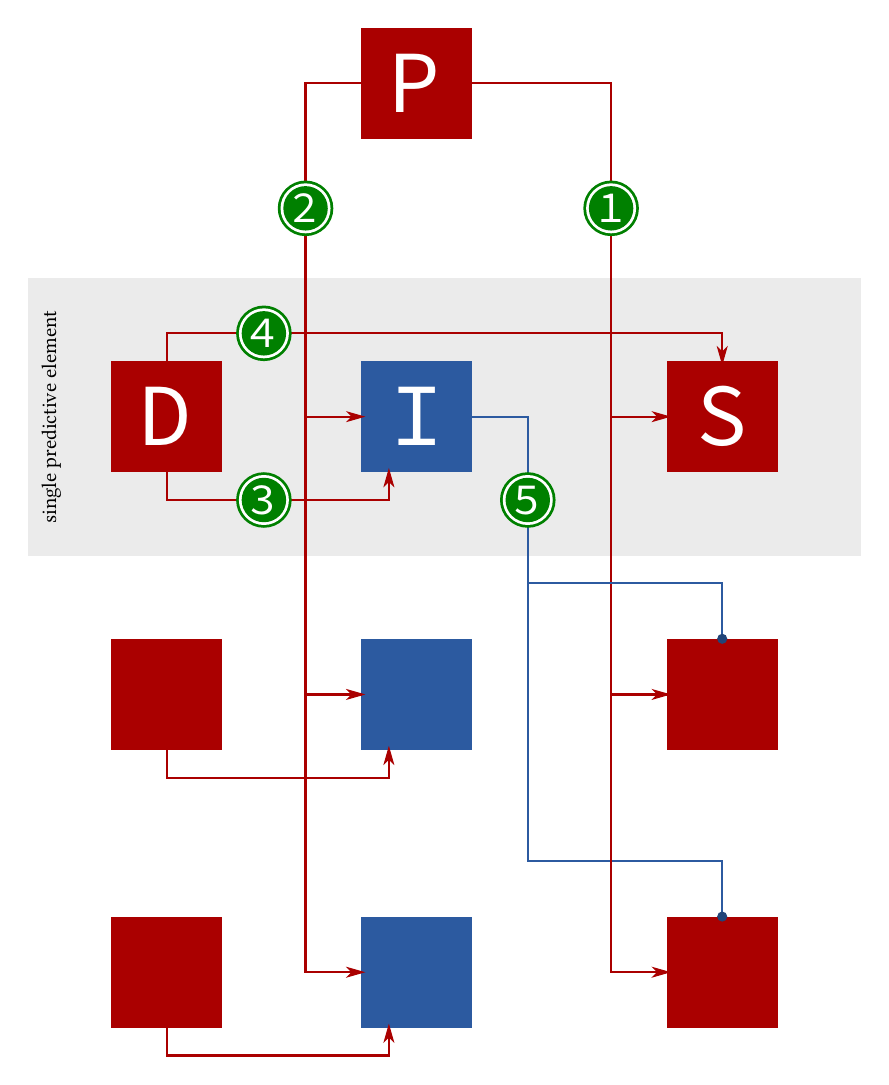}
	\end{center}
	\caption{Implementation of the temporal memory not including plasticity. Every HTM cell within a column is modeled with three individual LIF cells modeling different compartments (distal dendrites \emph{D}, soma \emph{S} and a lateral inhibition segment \emph{I} -- which is not biologically inspired). Per column, there exist multiple cell triples as well as one ``head'' cell \emph{P} which participates in the columnar competition and collects proximal input for the whole column. Activity of this cell is forwarded to the individual soma cells of the column \protect\circled{green}{1}. Without a previous prediction, this results in all soma cells firing. However, the distal compartment sums over the input of the previous time step. When a threshold is reached, the inhibitory compartment as well as the soma are depolarized \protect\circled{green}{3} \protect\circled{green}{4}. Together with proximal input \protect\circled{green}{2}, the inhibitory partition fires and inhibits all other cells in the column \protect\circled{green}{5}.}
	\label{fig:temporal_memory}
\end{figure}

Similar to the spatial pooler, the temporal memory implementation was  designed
for fast reaction times and spike-timing based response patterns. A complete
network consists of $m$ identical columns with $n$ \gls{htm} cells each.
Modelling these cells is a challenge in itself. A multicompartmental neuron
model would represent the best fit. While a neuromorphic hardware chip
implementing such a model is planned and first steps in that direction have
already been taken \citep{millner2012development}, the current system does not
provide this feature. Since \gls{htm} cells primarily depend on the active
properties of a compartment, it can be modelled by a triple of individual
\gls{lif} cells as shown in Figure~\ref{fig:temporal_memory}.

A column collects proximal input using a single cell. In fact, this cell can be
part of a spatial pooler network as presented in
section~\ref{sss:spatial_pooler_network}. When the column becomes active, this
cell emits a spike and excites both the neurons representing the \gls{htm}
cells' somae as well as inhibitory cells. The inhibitory projection, however, is
not strong enough to activate the target compartment alone. Instead, it only
leads to a partial depolarization. The soma neuron, however, reaches the firing
threshold for a single presynaptic event. This suffices as a columnar bursting
mechanism (i.e. temporal memory property 3): without predictive input, all soma
compartments will fire as a response to the proximal stimulus.

Distal input is processed for each cell individually by their dendritic segment
compartments. A cell's dendritic segment receives input from other cells' somae.
When its firing threshold is crossed, it partly depolarizes the inhibitory
helper cell of the same triplet. This synaptic projection is set up with a
relatively long synaptic time constant and a reversal potential matching the
threshold voltage. This ensures that the predictive state is carried to the next
time step and prohibits the cell from becoming active due to distal input alone.
On proximal input, the already depolarized helper cell fires before the somatic
compartments. The latter are then inhibited instantly, with the exception of the
own triplet's soma. As described, this basic predictive mechanism fails when
multiple cells are predicted, since the inhibitory compartments laterally
inhibit every cell. The solution is to also depolarize the somatic
compartments of predicted cells. In summary this mechanism satisfies temporal
memory properties 1 and 2.

\begin{figure*}[p]
	\begin{center}
		\input{assets/temporal_memory/temporal_memory_traces.pgf}
	\end{center}
	\caption{Neuron traces for a temporal memory column containing three \gls{htm} cells. Each of these cells is represented by a somatic compartment, an inhibitory helper cells and two dendritic segments. The column is activated by proximal input in every time step and receives random distal stimulus predicting none, one or more cells per step. As indicated by the automatic classification algorithm, the column exhibits a correct response pattern to these predictions.}
	\label{fig:static_temporal_memory_traces}
\end{figure*}

\section{Results}
The network models presented in the previous section were simulated in software to investigate their behavior. In the following, respective experiments and their results are shown. Additionally, plasticity rules and topological requirements are discussed in respect of the \gls{hmf}.

\subsection{Network Simulations}


The spatial pooler was analyzed for a network spanning 1,000 columns and an
input vector of size 10,000. To speed up the simulation, the input connectivity
was preprocessed in software by multiplying the stimulus vector to the
connectivity matrix.

\begin{figure}
	\begin{center}
		\input{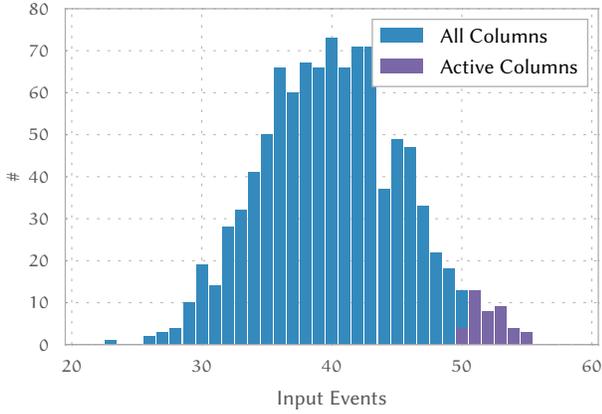}
	\end{center}
	\caption{Histogram showing the distribution of overlap scores individual columns receive. Columns activated by the spatial pooler network are highlighted. This confirms that only competitors with the highest input enter an active state. Furthermore, tie situations between columns with the same overlap score are resolved correctly.}
	\label{fig:spatial_pooler_activity}
\end{figure}

A first experiment was designed to verify the basic \gls{kwta} functionality. A
random pattern was presented to the network. The number of active inputs per
column -- the input overlap score -- can be visualized in a histogram as shown
in Figure~\ref{fig:spatial_pooler_activity}. By highlighting the columns
activated by that specific stimulus, one can investigate the network's selection
criteria. Complying with the requirements for a spatial pooler, only the
rightmost bars -- representing columns with the highest input counts -- are
highlighted. Furthermore, the model's capability to resolve ties between columns
receiving the same input counts is demonstrated: the bar at the decision
boundary was not selected as a whole but only a few columns were picked. This
verifies spatial pooler property~\ref{enm:spatial_pooler_selection}.

\begin{figure}
	\begin{center}
		\input{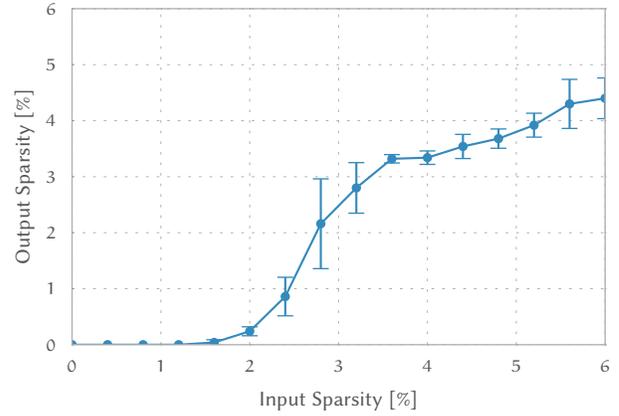}
	\end{center}
	\caption{The relative number of active columns is plotted against the input vector's sparsity. After a certain level of input sparsity is reached, columns start to enter active states. With higher presynaptic activity, columnar competition increases and the output sparsity reaches a plateu. The curve's exact course can be manipulated through the neurons' parameters as can the size of the plateau. Error bars indicate the standard deviation across five trials.} 
	\label{fig:spatial_pooler_sparsity}
\end{figure}

In a second scenario, input vectors with varying sparsity were fed into the
network, as shown in Figure~\ref{fig:spatial_pooler_sparsity}. The number of
active columns stays approximately constant across a wide range of input
sparsity. Additionally the plot shows that columns must receive a minimum amount
of input to become active at all. This verifies the underlaying \gls{kwta}
approach as well as spatial pooler properties~\ref{enm:spatial_pooler_sparsity}
and~\ref{enm:spatial_pooler_minimum}.

\begin{figure}
	\begin{center}
		\input{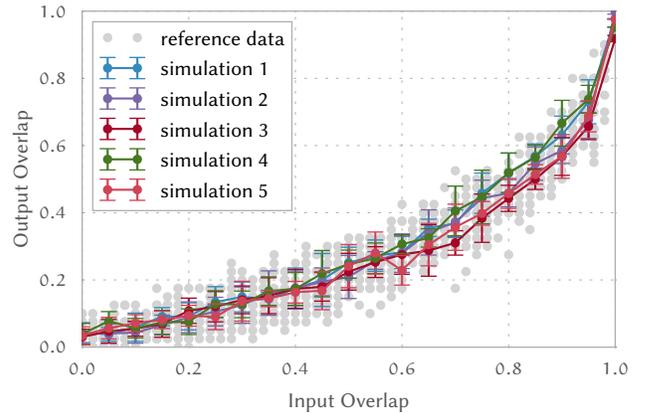}
	\end{center}
	\caption{Output overlap as a dependency of the input vector's overlap score. In each of the five simulation runs, the stimulus' was gradually changed starting from a random vector. As required for a spatial pooler, two similar input stimuli get mapped to similar output patterns, while disjunct input vectors result in low overlap scores. The simulations fully reproduce data from an existing software implementation which is also shown in this figure.}
	\label{fig:spatial_pooler_overlap}
\end{figure}

To verify the general functionality of a spatial pooler,  expressed in
property~\ref{enm:spatial_pooler_overlap}, a third experiment was set up. Input
data sets with a variable overlap were generated starting from an initial random
binary vector. For each stimulus, the overlap of the columnar activity with the
initial dataset was calculated while sweeping the input's overlap. The resulting
relation of input and output overlap scores is shown in
Figure~\ref{fig:spatial_pooler_overlap}. Also included are the results of a
similar experiment performed with a custom Python implementation of the spatial
pooler directly following the original specification \citep{numenta2011htm}.
Multiple simulation runs all yielded results perfectly matching the reference
data, thus verifying property~\ref{enm:spatial_pooler_overlap}.

The experiments have shown that the model presented in this section does fulfill
the requirements for a spatial pooler and can be considered a solid \gls{kwta}
implementation. The specific results of course depend on the individual
network size and configuration. In this case, the network -- most importantly
the columnar neurons' time constants -- was configured for a relatively short
time step of $T = \SI{50}{\milli\second}$. By choosing different parameter sets,
the network can be tuned towards different operational scenarios, e.g. further
increasing the model's stability.


The temporal memory was verified in a first sequence prediction
experiment. A reference software implementation was trained with three disjunct
sequences of length three. Consecutive sequences were separated by a random
input pattern. The trained network's lateral connectivity was dumped and loaded
in a simulation. When presented with the same stimulus, the \gls{lif}-based
implementation was able to correctly predict sequences, as shown in
Figure~\ref{fig:static_temporal_memory_live}.

\begin{figure*}[p]
	\begin{center}
		\input{assets/temporal_memory/live.pgf}
	\end{center}
	\caption{A \gls{lif} neuron based temporal memory implementation correctly predicting different patterns. Predicted cells are marked blue, active cells in purple. The network spans 128 columns with each of their eight \gls{htm} cells collecting distal stimuli via two dendritic segments. Connectivity for the distal inputs was configured externally. The model was presented three disjunct sequences of size three. The individual patterns were separated by a random input \gls{sdr}.}
	\label{fig:static_temporal_memory_live}
\end{figure*}

\subsection{Learning Algorithms}

Implementing online learning mechanisms in neuromorphic hardware is a challenge,
especially for accelerated systems. Although the \gls{hmf} features
implementations of nearest-neighbor \gls{stdp} and \gls{stp}
\citep{friedmann13plasticity,billaudelle14stp}, more complex update algorithms
are hard to implement. Numenta's networks rely on structural
plasticity rules which go beyond these mechanisms.

\begin{figure}
	\begin{center}
		\input{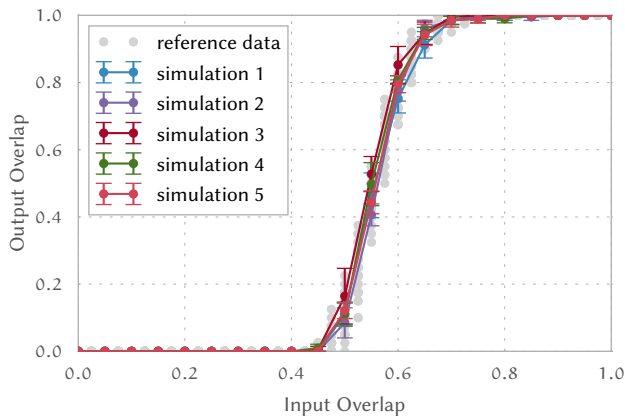}
	\end{center}
	\caption{Dependency of output and input overlap for a trained spatial pooler. Results of five independent simulation runs are shown as well as reference data from a custom software implementation.}
	\label{fig:spatial_pooler_learning}
\end{figure}

The spatial pooler's stimulus changes significantly for learned input patterns.
Verification of its functionality under these conditions is important. In order
to follow the \gls{htm} specification as closely possible, a supervised update
rule was implemented in an outer loop: for each time step, a matrix containing
the connections' permanence values is updated according to the activity patterns
of the previous time step. This allows us to implement the concepts of
structural plasticity presented in the original whitepaper. For the target
platform, the learning algorithms could be implemented on the \gls{ppu} which is
planned for the next version of the \gls{hicann} chip \citep{friedmann2013phd}.
Simulation results of the implementation described above are shown in
Figure~\ref{fig:spatial_pooler_learning}.

Experiments to replace the \gls{htm} structural plasticity rules by a classic
nearest-neighbor \gls{stdp} model did not yield the desired results. The
\gls{htm} learning rules require negative modifications to inactive synapses in
segments that contribute to cell activity. In contrast, \gls{stdp} does not
affect inactive synapses.

\subsection{Map and Route}

Applying abstract network models to the hardware platform requires algorithms for placing the neuron populations and routing the synaptic connections. In a best-case scenario, this processing step results in an isomorphic projection of the network graph to the hardware topology. For networks with extreme connectivity requirements, however, synaptic losses must be expected.

Mapping the simulated networks does not represent a challenge for the routing algorithms. The temporal memory can be projected to a single wafer without synaptic loss. The same still applies with assumed lateral all-to-all connectivity resulting in approximately 2 million synapses. The latter network corresponds to a network with a potential pool of \SI{100}{\%} which would allow the exploration of learning algorithms even without creating and pruning hardware synapses.

On the hardware platform, a tradeoff between the number of afferent connections
per cell and the number of neurons must be taken into consideration: while it is
possible to connect the dendritic membrane circuits such that a single neuron
can listen on roughly \num{14e3} synases, such a network could only consist of
approximately \num{3e3} neurons per wafer. With just 226 synapses, just under
\num{200e3} neurons can be allocated per wafer.

Scaling up the proof-of-concept models to a size useful for production purposes, however, challenges the hardware topology as well as the projection algorithms.

A minimal useful \gls{htm} network spans 1024 columns with 8 cells each. In such
a scenario the neurons would receive lateral input on 32 dendritic segments.
Allowing approximately \num{1e3} afferent connections per dendritic segment,
this network could be realized on approximately \num{1e6} dendritic membrane
circuits, or six wafers. The existing system set up for the \gls{bss} would
suffice for this scenario. Even larger networks could be brought to the
\gls{hbp}'s platform.

\section{Conclusion and Outlook}
Implementing machine intelligence algorithms as spiking neural networks and
porting them to a neuromorphic hardware platform presents high demands in terms
of precision and scalability.

We have shown in this paper that \glspl{htm} can be successfully modeled in
dynamic simulations. The basic functionality of spatial pooler and temporal
memory networks could be reproduced based on \gls{adex} neurons. In theory, the
proof of concept networks can be easily transferred to the \gls{hmf}, since the
high-level software interfaces are designed to be interchangable. Of course,
emulating the models on the actual hardware platform will bring up a new set of
challenges.

Adapting the \gls{htm}'s learning rules to the native plasticity features
available on the \gls{hmf} has turned out to be nontrivial. The learning rules
could not be replicated with the current implementation of classic \gls{stdp}.
As a freely programmable microprocessor directly embedded into the neuromorphic
core, the \gls{ppu} provides the ability to extend the system's plasticity
mechanisms in order to implement the \gls{htm} rules. Further investigation is
required to map out a complete implementation of the \gls{htm} update rules on
the \gls{ppu}.

Analog neuromorphic hardware is susceptible to transistor mismatches due to e.g. dopand fluctuations in the production process \citep{petrovici2014characterization}. A careful calibration of the individual neurons is required to compensate for these variations. Due to the complexity of the problem and the high number of interdependent variables, a perfect calibration is hard to accomplish. Therefore, network models are required to be tolerant regarding certain spatial, and trial-to-trial variations on the computing substrate. Carrying out additional Monte Carlo simulations with slightly randomized parameters is important to investigate the robustness of the presented networks.

Finally, a multicompartmental neuron model is planned for a later version of the neurmorphic platform. Making use of this extended feature set will significantly increase the level of biophysical detail. This will account for the detailed dendritic model used in \glspl{htm} and help to stay closer to the whitepaper as well as the reference implementation.

Besides paving the road towards a highly accelerated execution of \gls{htm}
models, the \gls{hmf} offers a high level of detail in its neuron
implementation. With the multicompartmental extension and a flexible plasticity
framework, we anticipate the platform will prove valuable as a tool for
further low-level research on \gls{htm} theories.

\section*{Acknowledgements}

Special thanks to Jeff Hawkins, Prof. Dr. Karlheinz Meier, Paxon Frady,
and the Numenta Team.

\bibliography{report}

%
%
%
%
%
%
%
\end{document}